# LEP- AND LOW-ENERGY BOUNDS ON $R$-PARITY-VIOLATING SUPERSYMMETRIC YUKAWA COUPLINGS

GAUTAM BHATTACHARYYA

*Theory Division, CERN, CH–1211 Geneva 23, Switzerland.*

## 1 Introduction

'$R$-parity', equivalently known as 'matter parity', correponds in supersymmetry (SUSY) to a discrete symmetry following from the conservation of lepton-number ($L$) and baryon-number ($B$)[1]. It is represented as $R = (-1)^{(3B+L+2S)}$, where $S$ is the intrinsic spin of the field. $R$ is $+1$ for all standard model (SM) particles and $-1$ for all super-particles. However, $B$- and $L$-conservations are not ensured by gauge invariance. This makes the issue of putting phenomenological bounds on the strengths of the $L$- and $B$-violating, or more generally the $R$-parity-violating, supersymmetric Yukawa couplings a potentially interesting one. SUSY requires the presence of two Higgs doublets and the gauge quantum numbers of one of the two Higgs super-multiplets are the same as those of the $SU(2)$-doublet leptonic superfield. So the latter can replace the former in the Yukawa interaction terms, if one sacrifices the requirement of $L$-conservation. One can also write $B$-violating Yukawa interaction involving three $SU(2)$-singlet quark superfields. These lead to *explicit* breaking of $R$-conserving interactions, which can be parametrized as

$$\mathcal{W}_{\not{R}} = \lambda_{ijk} L_i L_j E_k^c + \lambda'_{ijk} L_i Q_j D_k^c + \lambda''_{ijk} U_i^c D_j^c D_k^c \ , \quad (1)$$

where $L_i$ and $Q_i$ are the $SU(2)$-doublet lepton and quark superfields and $E_i^c, U_i^c, D_i^c$ are the singlet superfields; $\lambda_{ijk}$ is antisymmetric under the interchange of the first two $SU(2)$ indices, while $\lambda''_{ijk}$ is antisymmetric under the interchange of the last two. This means that there are 27 $\lambda'$-type and 9 each of $\lambda$- and $\lambda''$-type couplings, thus adding 45 extra parameters in the minimal SUSY. It may be noted that $\lambda$- and $\lambda'$-types are $L$-violating, while $\lambda''$-types are $B$-violating couplings. A consequence of the presence of the above terms is that the lightest supersymmetric particle (LSP) is not stable, which makes the SUSY search strategies different from what they would have been in the absence of $R$-violating interactions.

## 2 Cosmological implications

There exist important cosmological constraints[2] on $R$-parity-violating scenarios. The requirement that GUT-scale baryogenesis does not get washed out imposes $\lambda'' \ll 10^{-7}$, although these bounds are model dependent and can be evaded[3]. If the $B$-violating couplings are not present, the $\lambda'$ couplings cannot wash out the initial baryon asymmetry by themselves. However, they can do so in association with a $B$-violating but $(B-L)$ conserving interaction, such as sphaleron-induced non-perturbative transitions. It may be noted that these processes conserve $\frac{1}{3}B - L_i$ for each lepton generation, and hence the conservation of any one lepton generation number is sufficient to retain the initial baryon asymmetry. Therefore, the assumption that the smallest $\lambda'$-type coupling is less than $\sim 10^{-7}$ is enough to avoid any cosmological bound on the remaining $\lambda'$-type couplings.

## 3 Phenomenological studies

There have been extensive phenomenological studies for putting bounds on the $R$-parity-violating couplings from low-energy processes. Here we mention only a few. The simultaneous presence of the $\lambda'$- and $\lambda''$-type couplings is very strongly constrained ($\lambda', \lambda'' \leq 10^{-10}$) from non-observation of proton decay. The most serious constraints on the $B$-violating couplings originate from the absence of $n$–$\bar{n}$ oscillation and the correponding heavy nuclei decay[4], yielding $\lambda''_{112} \leq 10^{-8}$ for $m_{\tilde{q}} = 100$ GeV. The strongest constraints on the $L$-violating couplings follow from the upper limit of the $\nu_e$-Majorana mass[5] imposing $\lambda_{133} \leq 3 \times 10^{-3}$ and $\lambda'_{133} \leq 10^{-3}$ for the same squark mass as mentioned above. Besides, there are bounds on other $\lambda$-and $\lambda'$-types couplings from charged-current universality, $e$-$\mu$-$\tau$ universality, $\nu_\mu$-$e$ scattering, atomic parity violation, etc. A list of these limits can be found in Table 1 of Barger et al.[6]. Bounds on the product couplings of the $\lambda''$-type have been recently reported[7].

It may be noted that most of the Yukawa couplings that involve the third family are not constrained from low-energy processes. We attempt to put new bounds to many of these 'yet unconstrained' couplings and also improve the limits on some others. We use (i) the LEP electroweak observables to put bounds[8,9] on $\lambda'_{i3k}$ for all $i$ and $k$ and on $\lambda''_{3jk}$ for all possible $j$ and $k$, and (ii) the experimental data on $D$-decays to constrain $\lambda'_{12k}$ and $\lambda'_{22k}$ as well as the data on $\tau$-decay to put bounds[10] on $\lambda'_{31k}$ (for all $k$). These are briefly outlined below:

## 3.1 LEP electroweak observables

The decays of $Z$ are in general very sensitive to the third-family-induced vertex corrections. The $\lambda'$-induced vertex corrections to $R_l = \Gamma_{\text{had}}/\Gamma_l$ can have a sizeable contribution. There are new triangle diagrams contributing to $\Gamma_l$ with $Z, l^+$ and $l^-$ external lines involving $\lambda'_{ijk}$ vertices with $i$ = lepton, $j$ = quark, $k$ = squark indices or $i$ = lepton, $j$ = squark, $k$ = quark indices. Such couplings can also affect $\Gamma_{\text{had}}$ through triangle diagrams where the external lines are $Z, q$ and $\bar{q}$ in a situation where, for example, $i$ = slepton, $j$ = quark (squark) and $k$ = squark (quark). Since the magnitude of the new contribution *essentially* depends on the mass of the fermion in the loop, only $\lambda'_{i3k}$-type couplings leading to internal top quark lines can be constrained significantly by our considerations. Similarly, the $\lambda''$-induced vertex corrections to the decay widths $Z \to q\bar{q}$ can affect $R_l$ through its numerator, and the top-quark induced corrections again turn out to be significant. The bounds we obtain are listed in Table 1 of reference 8 and Table 2 of reference 9. In short, for $m_{\tilde{q}} = 100$ GeV and at $1\sigma$, the following bounds emerge:[a]

$$\begin{aligned}
\lambda'_{13k} &\leq 0.51 \leftarrow R^{\text{exp}}_e = 20.850 \pm 0.067, \\
\lambda'_{23k} &\leq 0.44 \leftarrow R^{\text{exp}}_\mu = 20.824 \pm 0.059, \\
\lambda'_{33k} &\leq 0.26 \leftarrow R^{\text{exp}}_\tau = 20.749 \pm 0.070, \quad (2) \\
\lambda''_{3jk} &\leq 0.97 \leftarrow R^{\text{exp}}_l = 20.795 \pm 0.040.
\end{aligned}$$

The above experimental data are taken from the LEP Electroweak Working Group report[11].

## 3.2 D- and τ-decays

We use the following experimental inputs[12]:

$$\begin{aligned}
a) & \quad \frac{Br(D^+ \to \bar{K}^0 \mu^+ \nu_\mu)}{Br(D^+ \to \bar{K}^0 e^+ \nu_e)} = 1.06^{+0.48}_{-0.34}; \\
b) & \quad \frac{Br(D^+ \to \bar{K}^{0*} \mu^+ \nu_\mu)}{Br(D^+ \to \bar{K}^{0*} e^+ \nu_e)} = 0.94 \pm 0.16; \quad (3) \\
c) & \quad \frac{Br(D^0 \to K^- \mu^+ \nu_\mu)}{Br(D^+ \to K^- e^+ \nu_e)} = 0.84 \pm 0.12;
\end{aligned}$$

and

$$\begin{aligned}
d) \quad Br(\tau^- \to \pi^- \nu_\tau) &= 0.117 \pm 0.004, \quad (4) \\
f_\pi &= (130.7 \pm 0.1 \pm 0.36) \text{ MeV}.
\end{aligned}$$

In the case of $D$-decays the form factors associated with the hadronic matrix elements cancel in the ratios, thus making our predictions free from the large theoretical uncertainties associated with them. The bounds we obtain are listed in Table 1 of reference 10. A summary of the table, for $m_{\tilde{q}} = 100$ GeV and at $1\sigma$, is

$$\begin{aligned}
\lambda'_{12k} &\leq 0.29 \quad \text{from (b)}, \\
\lambda'_{22k} &\leq 0.18 \quad \text{from (b)}, \quad (5) \\
\lambda'_{31k} &\leq 0.16 \quad \text{from (d)}.
\end{aligned}$$

## 4 Conclusion

*Many of our bounds are new.* From $Z$-physics at LEP we have obtained bounds on 9 $\lambda'$-type (7 are new bounds) and 3 $\lambda''$-type couplings (all are new bounds). Of the bounds obtained from $D$-decays, 2 are new and the rest are at par with the existing ones, while the ones obtained from $\tau$-decay are all new.

## Acknowledgments

I thank D. Choudhury, J. Ellis and K. Sridhar for sharing their insights during a few enjoyable collaborations on which this article is based.

---

[a] While extracting limits on $\lambda''$, leptonic universality is assumed since they do not couple to any leptonic flavour.